\RequirePackage{fix-cm}

\documentclass[%
a4paper, %
intlimits, %
twocolumn, %
]{revtex4-1}
\usepackage{cmap}

\usepackage{ucs}
\usepackage[utf8x]{inputenc}
\usepackage[T2A]{fontenc}
\usepackage[russian,english]{babel}

\usepackage{amsmath}
\usepackage{amssymb}

\usepackage{mathtools}
\mathtoolsset{
showonlyrefs,
mathic = true
}

\allowdisplaybreaks

\usepackage{hyperref}
\hypersetup{backref,
 colorlinks=false}
\hypersetup{pdfborder=0 0 0}
\usepackage{url}

\usepackage{nicefrac}

\usepackage{physymb}

\makeatletter
\def\ps@pprintTitle{%
     \let\@oddhead\@empty
     \let\@evenhead\@empty
     \let\@oddfoot\@empty
     \let\@evenfoot\@oddfoot}
\makeatother

\newcommand{\e}{\mathrm{e}}
\renewcommand{\i}{\mathrm{i}}

\newcommand{\setR}{\mathbb{R}}
\newcommand{\setS}{\mathbb{S}}

\newcommand{\crd}[1]{\underline{\vphantom{j}{#1}}}
\begin{document}

\author{D. S. Kulyabov}
\email{yamadharma@gmail.com}
\author{A. G. Ulyanova}
\email{jelly26a@gmail.com}

\affiliation{Peoples' Friendship University of Russia}

\thanks{Published in: D.~S. Kulyabov and A.~G. Ul’yanova.  Application
  of two-spinor calculus in quantum mechanical and field calculations.
  \emph{Physics of Particles and Nuclei Letters}, 6 (7): 546--549,
  2009.
  \href{http://dx.doi.org/10.1134/S1547477109070115}{doi:~10.1134/S1547477109070115}.
  URL \url{http://www.springerlink.com/content/bhmw776v15110271/}.}

\thanks{Sources:\\
  \url{https://bitbucket.org/yamadharma/articles-2008-two-spinors}}

\title{An application of two-spinors calculus to quantum field and
  quantum mechanics problems}

\begin{abstract}
  This paper describes the Lorentz two-spinors proposing to use them
  instead of Dirac four-spinors and quaternions.
\end{abstract}

\maketitle

  \section{Introduction}
\label{sec:intro}

  Spinors are used in physics quite extensively~\cite{cartan2012theory}.  The
  following spinors are mainly used:
  \begin{itemize}
  \item Dirac four-spinors;
  \item Pauli three-spinors;
  \item quaternions.
  \end{itemize}

  If Dirac four-spinors are used, the main difficulty is $\gamma$
  matrices. The essence of these objects is that they serve to connect
  the spinor and tensor spaces and therefore have two types of
  indices: spinor and tensor ones.  It would be logical to perform
  calculations in one of these spaces only.

  We propose using semispinors of Dirac spinors, Lorentz
  two-spinors~\cite{penrose1987spinors}, as simpler objects.

  \section{General Notion of Spinors}
\label{sec:gener-noti-spin}

Let us determine a spinor using the Clifford–Dirac
equation:
\begin{equation}
  \label{eq:1}
  \gamma_{(a}\gamma_{b)}=-g_{ab}\mathbf{I}.
\end{equation}

Or omitting the spinor indices:
\begin{equation}
  \label{eq:2}
  \gamma_{a\rho}^{\sigma} \gamma_{b\sigma}^{\tau} +
  \gamma_{b\rho}^{\sigma} \gamma_{a\sigma}^{\tau}= -2 g_{ab}\delta_{\rho}^{\tau}.
\end{equation}

The dimension of the spinor space is:
\begin{equation}
  \label{eq:3}
    \begin{cases}
      N = 2^{n/2}, & \text{even } n; \\
      N = 2^{n/2 -1/2}, & \text{odd } n.
    \end{cases}
\end{equation}

  \section{Quaternions and Two-Spinors}
\label{sec:quat-two-spin}

  Let us assume that
\begin{equation}
  \begin{gathered}
    \label{eq:quat.1}
    \mathbf{I} =
    \begin{pmatrix}
      1 & 0 \\
      0 & 1
    \end{pmatrix}, \quad \mathbf{i} =
    \begin{pmatrix}
      0 & \i \\
      \i & 0
    \end{pmatrix}, \\
    \mathbf{j} =
    \begin{pmatrix}
      0 & -1 \\
      1 & 0
    \end{pmatrix}, \quad \mathbf{k} =
    \begin{pmatrix}
      \i & 0 \\
      0 & -\i
    \end{pmatrix}.
  \end{gathered}
\end{equation}

  A general quaternion is represented by the matrix
\begin{equation}
  \label{eq:quat.2}
  \mathbf{A} = \mathbf{I} a + \mathbf{i} b + \mathbf{j} c + \mathbf{j} d =
  \begin{pmatrix}
    a + \i d & -c + \i b \\
    c + \i b & a - \i d
  \end{pmatrix},
\end{equation}
  where $a,b,c,d \in \setR$.
  The sum and the product of two quaternions are obtained as the
  matrix sum and matrix product. The adjoint quaternion
  $\mathbf{A}^{*}$ is determined by the matrix operation:
\begin{equation}
  \label{eq:quat.3}
  \mathbf{A}^{*} = \mathbf{I} a - (\mathbf{i} b + \mathbf{j} c + \mathbf{k} d).
\end{equation}

  If matrix $\mathbf{A}$ \eqref{eq:quat.2} is unimodular and unitary,
  it represents the unitary spin matrix. The following relations can
  be written:
\begin{gather}
  \label{eq:quat.4}
  \det \mathbf{A} = a^2 + b^2 + c^2 +d^2 = 1, \\
  \label{eq:quat.5}
  \mathbf{A}\mathbf{A}^{*} = \mathbf{I} (a^2 + b^2 +c^2 +d^2) =
  \mathbf{I}.
\end{gather}

  Thus, the quaternion should have the unitary norm:
\begin{equation}
  \label{eq:quat.6}
  \| \mathbf{A} \| := a^2 + b^2 + c^2 +d^2 = 1.
\end{equation}
  Therefore, the unitary quaternion can be represented as the unitary
  spin matrix.

  In spite of the fact that unitary spin matrices and unitary
  quaternions represent in essence the same thing, in the general
  case, no close connection exists between quaternions and spin
  matrices. The fact of the matter is that quaternions are connected
  with positive definite quadratic forms, while spin matrices and
  Lorentz transforms are characterized by the Lorentz signature
  $(+,-,-,-)$.

\section{From Two-Spinors to Vectors}
\label{sec:from-spinors-vectors}

In final calculations, it is necessary to transform
abstract indices into component form. Moreover, it is
often more convenient to formulate the result in vector
form. To establish the connection between the spinor
and vector bases, the Infeld–van der Waerden symbols
are used:
\begin{equation}
  \label{eq:spinor.verden.1}
  \begin{gathered}
    g_{\crd{a}}{}^{\crd{AA'}} := g_{\crd{a}}{}^a
    \varepsilon_A{}^{\crd{A}}
    \varepsilon_{A'}{}^{\crd{A'}}, \\
    g^{\crd{a}}{}_{\crd{AA'}} := g_a{}^{\crd{a}}
    \varepsilon_{\crd{A}}{}^A \varepsilon_{\crd{A'}}{}^{A'},
  \end{gathered}
\end{equation}
where convolution is performed over abstract indices
only.

For the standard Minkowski tetrad and the spin ref-
erence frame, we obtain
\begin{equation}
  \label{eq:spinor.verden.2}
  \begin{gathered}
    g_0{}^{\crd{AB'}} = \frac{1}{\sqrt{2}}
    \begin{pmatrix}
      1 & 0 \\
      0 & 1
    \end{pmatrix} = g_{\crd{AB'}}^0, \\ 
    g_1{}^{\crd{AB'}} =
    \frac{1}{\sqrt{2}}
    \begin{pmatrix}
      0 & 1 \\
      1 & 0
    \end{pmatrix} = g_{\crd{AB'}}^1, \\
    g_2{}^{\crd{AB'}} = \frac{1}{\sqrt{2}}
    \begin{pmatrix}
      0 & \i \\
      -\i & 0
    \end{pmatrix} = - g_{\crd{AB'}}^2, \\
    g_3{}^{\crd{AB'}} =
    \frac{1}{\sqrt{2}}
    \begin{pmatrix}
      1 & 0 \\
      0 & -1
    \end{pmatrix} = g_{\crd{AB'}}^3.
  \end{gathered}
\end{equation}

\section{Dirac Four-Spinors and Lorentz Two-Spinors}
\label{sec:dirac-four-spinors}

  In physical calculations, Dirac four-spinors\footnote{ The
    introduction of four-spinors can probably be motivated by the
    desire to construct an object for which the operation of spatial
    reflection can be implemented conveniently.  } are often
  used. However, operations with these objects are extremely
  cumbersome. One of the basic disadvantages of this formalism is the
  explicit application of $\gamma$ matrices which are in essence
  objects serving to connect the vector and spinor spaces. It is as
  though our spinor objects ``live'' in two spaces, vector and spinor
  spaces.  Correspondingly, the desire to transfer to a more unified
  formalism by rejecting either spinor or tensor indices is observed.

  Let us consider the transition to the purely spinor formalism based
  on Lorentz two-spinors. First, we will construct a four-spinor
  object based on two-spinors.  Then we will demonstrate the
  capabilities of this formalism. Two examples will be considered: the
  derivation of invariant spinor relations and the calculation of
  matrix elements.

  \subsection{Construction of Four-Spinor Formalism}
\label{sec:qed.4spinor.build}

  Let us construct the implementation of the four-spinor formalism
  based on Lorentz two-spinors.\footnote{The Dirac four-spinor is
    often implemented using two three-dimensional spinors (three Pauli
    spinors). Formally, this is quite possible, since the structure of
    the spaces $\setS_{R}$ and $\setS_{R'}$ of the semispinors of the
    spinor with a dimension of $n+1$ coincides with the structure of
    the space $S_\rho$ of spinors with a dimension of $n$ ($n$ is
    odd).}

  We denote by small Greek letters four-spinor indices; by capital
  Latin letters, two-spinor indices (as usual); and by small Latin
  letters, tensor indices.

  Let us write the Dirac four-spinor as
\begin{equation}
  \label{eq:qed.4spinor.1}
  \psi^\alpha =
  \begin{pmatrix}
    \varphi^A \\
    \pi^{A'}
  \end{pmatrix},
\end{equation}
where $\varphi^A$ and $\pi^{A'}$ are Lorentz two-spinors.

  The adjoint spinor has the form
\begin{equation}
  \label{eq:qed.4spinor.2}
  \overline{\psi^\alpha} = \bar{\psi}_\alpha =
  \begin{pmatrix}
    \bar \pi_A , \bar \varphi_{A'}
  \end{pmatrix}.
\end{equation}

  Let us determine the reflection operator:
\begin{equation}
  \label{eq:qed.4spinor.3}
  \hat P =
  \begin{pmatrix}
    \varphi^A \\
    \pi^{A'}
  \end{pmatrix} \mapsto
  \begin{pmatrix}
    \pi^{A'} \\
    \varphi^{A}
  \end{pmatrix}.
\end{equation}

  We use $\gamma$ matrices in the chiral representation. Then the
  explicit form of the $\gamma$ matrix is
\begin{equation}
  \label{eq:qed.4spinor.4}
  \begin{gathered}
    \gamma_{a\rho}{}^\sigma = \sqrt{2}
    \begin{pmatrix}
      0 & \varepsilon_{A'R'} \varepsilon_{A}{}^{S} \\
      \varepsilon_{AR} \varepsilon_{A'}{}^{S'} & 0
    \end{pmatrix}, \\ 
    \eta_{\rho}{}^\sigma =
    \begin{pmatrix}
      -\i \varepsilon_{R}{}^S & 0 \\
      0 & \i \varepsilon_{R'}{}^{S'}
    \end{pmatrix},
  \end{gathered}
\end{equation}
  and
\begin{equation}
  \label{eq:qed.4spinor.5}
  \gamma_{ab\rho}{}^\sigma =
  \begin{pmatrix}
    \varepsilon_{A'B'} \varepsilon_{R(A} \varepsilon_{B)}{}^S & 0 \\
    0 & \varepsilon_{AB} \varepsilon_{R'(A'} \varepsilon_{B')}{}^{S'}
  \end{pmatrix}.
\end{equation}

  We introduce the following notation: $\gamma_5 := \i \eta$.

  Using the structure of Dirac four-spinors determined by us, it is
  possible to construct invariant relations. We will operate with the
  pair spinor–adjoint four-spinor ($\psi$ and $\bar\psi$), and
  correspondingly, four two-spinors ($\varphi^A$, $\bar\varphi_{A'}$,
  $\pi^{A'}$, and $\bar\pi_A$).

  \subsection{Scalars}
\label{sec:scalars}

  The convolutions $\bar\pi_A \varphi^A$ and
  $\bar\varphi_{A'}\pi_{A'}$ have the meaning of scalars. Their sum
  behaves as a scalar, and the difference, as a pseudoscalar:
\begin{align}
  \label{eq:qed.4spinor.6}
  s &= \bar\pi_A \varphi^A + \bar\varphi_{A'}\pi^{A'}
  = \bar\psi_\alpha \psi^\alpha, \\
  \label{eq:qed.4spinor.7}
  p &= \i (\bar\pi_A \varphi^A - \bar\varphi_{A'}\pi^{A'}) = \i
  \bar\psi_\alpha \gamma_{5 \beta}{}^\alpha \psi^\beta.
\end{align}

  \subsection{Vectors}
\label{sec:vectors}

  The combinations $\bar{\pi}^A\pi^{A'}$ and
  $\varphi^A\bar{\varphi}^{A'}$ have the meaning of vectors. Their sum
  behaves as a vector, and the difference, as a pseudovector:
\begin{align}
  \label{eq:qed.4spinor.8}
  j^a &= \sqrt{2} (\bar{\pi}^A\pi^{A'} + \varphi^A\bar{\varphi}^{A'})
  =
  \bar{\psi}_\alpha \gamma^{a}_ \beta{}^\alpha \psi^\beta,\\
  \label{eq:qed.4spinor.9}
  \tilde{j}^a &= \sqrt{2} (\bar{\pi}^A\pi^{A'} -
  \varphi^A\bar{\varphi}^{A'}) = \bar{\psi}_\alpha
  \gamma^{a}_\beta{}^\alpha \gamma_{5 \delta}{}^\beta \psi^\delta.
\end{align}

  \subsection{Tensors}
\label{sec:tensors}

  The real antisymmetric tensor can be build in a following way:
\begin{equation}
  \label{eq:qed.4spinor.10}
  a^{ab} = \i (\varphi^{(A}\bar{\pi}^{B)} \varepsilon^{A'B'} -
  \bar{\varphi}^{(A'}\pi^{B')} \varepsilon^{AB}) = \bar{\psi}_\alpha
  \sigma^{ab}{}_\beta{}^\alpha \psi^\beta.
\end{equation}

  We think, this is more simple than with four-spinors.

  \section{Matrix Elements}
\label{sec:qed.matrix}

  Usually, the ``Feynman trick'' is used for calculating matrix
  elements in quantum theory; this trick consists in the
  transformation of the product of spinors into the spur; as a result,
  the squared matrix element is obtained.  Correspondingly, the
  complexity of calculations increases and the number of calculated
  elements is proportional to $n^2$. Moreover, if the complete matrix
  element is calculated as the sum of many diagrams, or the
  information on the phase is important, this method is inapplicable.

  The proposed alternative is to calculate the matrix element. Let us
  look at two ways: the application of two-spinors (rejection of
  tensor indices) and the application of the vector formalism
  (rejection of spinor indices).

  To eliminate the basic obstacle, complex relations for $\gamma$
  matrices, we propose using the two-spinor formalism.

  Let us introduce the auxiliary notation based on the sign in the
  projector $1 \pm \gamma_5$:
\begin{equation}
  \label{eq:qed.matrix.1}
  \psi =
  \begin{pmatrix}
    \psi_{+} \\
    \psi_{-}
  \end{pmatrix}.
\end{equation}

  Correspondingly, the $\gamma$ matrices are written in the form (see
  \eqref{eq:qed.4spinor.4}):
\begin{equation}
  \label{eq:qed.matrix.2}
  \gamma_a =
  \begin{pmatrix}
    0 & \gamma_{a\,+}\\
    \gamma_{a\,-} & 0
  \end{pmatrix}, \qquad \hat{p} =
  \begin{pmatrix}
    0 & \hat{p}_{+} \\
    \hat{p}_{-} & 0
  \end{pmatrix},
\end{equation}
  where $\hat{p}:= p^a \gamma_a$.

  The application of two-spinors is especially justified in the case
  of the presence of projectors $(1 \pm \gamma_5)$. Therefore, let us
  consider the simplification of calculations in the case of terms of
  the form
\begin{equation}
  \label{eq:qed.matrix.3}
  \bar{\psi}_f \gamma^{a_1} \hat{p}_{(a)} \gamma^{a_2} \hat{p}_{(b)}
  \cdots \gamma^{a_n} \left[ \frac{1}{2} (1 \pm \gamma_5) \right] \psi_i.
\end{equation}

  Two cases can be singled out: the even and odd numbers of $\gamma$
  matrices:
\begin{equation}
  \label{eq:qed.matrix.4}
  \begin{cases}
    \bar{\psi}_{f \pm} \gamma^{a_1}_{\mp} \hat{p}_{(a) \pm}
    \gamma^{a_2}_\mp \hat{p}_{(b) \pm} \cdots \gamma^{a_n}_\pm \psi_{i
      \pm}, \\
    \text{odd number of $\gamma$ matrix;}
    \\
    \bar{\psi}_{f \mp} \gamma^{a_1}_{\pm} \hat{p}_{(a) \mp}
    \gamma^{a_2}_\pm \hat{p}_{(b) \mp} \cdots \gamma^{a_n}_\mp \psi_{i
      \pm}, \\
    \text{even number of $\gamma$ matrix}.
  \end{cases}
\end{equation}

  Using the two-spinor representation of $\gamma$ matrices, we obtain
  the following relations:
\begin{subequations}
  \label{eq:qed.matrix.5}
  \begin{gather}
    \label{eq:qed.matrix.5a}
    \gamma_{a\alpha}{}^\beta_{+} \gamma_{\gamma}^a{}^\delta_{-} = 2
    \delta_\alpha{}^\delta \delta_\gamma{}^\beta, \\
    \label{eq:qed.matrix.5b}
    \gamma_{a \alpha}{}^\beta_{\pm} \gamma_{\gamma}^a{}^\delta_{\pm} =
    2 \left( \delta_\alpha{}^\beta \delta_\gamma{}^\delta -
      \delta_\alpha{}^\delta \delta_\gamma{}^\beta \right).
  \end{gather}
\end{subequations}

  For example, let us prove \eqref{eq:qed.matrix.5a}:
\begin{equation}
  \label{eq:64}
  \varepsilon_{C'A'} \varepsilon_C{}^B \varepsilon^C{}_G
  \varepsilon^{C'D'} = 2 \varepsilon_{A'}{}^{D'} \varepsilon_G{}^B,
\end{equation}

  Taking into account the following correspondences:
\begin{equation}
  \alpha \leftrightarrow A', \quad \beta \leftrightarrow B, \quad
  \gamma \leftrightarrow G, \quad \delta \leftrightarrow D',
\end{equation}
  we obtain the initial expression.

  Thus, sequentially using \eqref{eq:qed.matrix.4} and
  \eqref{eq:qed.matrix.5}, we eliminate $\gamma$ matrices and perform
  calculations using two-spinors.

  After calculations, terms of the following type are obtained:
\begin{equation}
  \label{eq:qed.matrix.6}
  u_{f\pm}^\dag \hat{p}_{(a)\mp} \hat{p}_{(b)\pm} \dots \hat{e}_{+\text{
      or }-} \dots u_{i\pm},
\end{equation}
  where $e$ is polarization.

  To obtain a particular result, it is necessary to choose the spinor
  representation. For example, the standard solution obtained by the
  decomposition into plane waves can be used. Then in the case of
  longitudinal polarization, we obtain
\begin{equation}
  \label{eq:qed.matrix.7}
  u_{\pm} = \left( \sqrt{E+\varepsilon m} \pm \varepsilon s \sqrt{E-
      \varepsilon m} \right)
  \begin{pmatrix}
    \e^{-\i \varphi/2} \sqrt{1 + s \cos \theta} \\
    \e^{\i \varphi/2} \sqrt{1 - s \cos \theta}
  \end{pmatrix},
\end{equation}
  where $s$ is the helicity and $\varepsilon$ is the energy sign.

  \subsection{Example of Calculation of a Matrix Element}
\label{sec:qed.matrix.example}

  Let us calculate the cross section of the reaction
\begin{equation}
  \label{eq:qed.matrix.example.1}
  \nu + n \to p + e^{-},
\end{equation}

  The matrix element of this reaction in the standard (V–A) theory has
  the form
\begin{equation}
  \label{eq:qed.matrix.example.2}
  M = \frac{G_F}{\sqrt{2}} \left( \bar{\psi}_e \gamma_a (1 + \gamma_5)
    \psi_\nu \right) \left( \bar{\psi}_p \gamma^a (g_V + g_A \gamma_5)
    \psi_n \right).
\end{equation}

  Using \eqref{eq:qed.matrix.4}, \eqref{eq:qed.matrix.5}, we reduce
  \eqref{eq:qed.matrix.example.2} to the following form:
\begin{multline}
  \label{eq:qed.matrix.example.3}
  M = \frac{2 G_F}{\sqrt{2}} \bigl(u^{\dag}_{e\alpha +} \gamma_{a
    \beta -}{}^\alpha u_{\nu +}^\beta \bigr) 
  \times {} \\ {} \times
  \biggl[ (g_A - g_V)
  \bigl( u^{\dag}_{p \gamma +} \gamma^a_{\delta -}{}^\gamma
  u^\delta_{n +} + u^{\dag}_{p \gamma -} \gamma^a_{\delta +}{}^\gamma
  u^\delta_{n -} \bigr) + {} \\
  {} + 2 g_A u^{\dag}_{p \gamma +} \gamma^a_{\delta -}{}^\gamma
  u^\delta_{n +} \biggr] 
  = {} \\ {} =
  \frac{2 G_F}{\sqrt{2}} \biggl[ (g_V - g_A)
  u^{\dag}_{e \alpha +} \gamma_{a \beta -}{}^\alpha u^\beta_{\nu +}
  u^{\dag}_{p \gamma -} \gamma^a_{\delta +}{}^\gamma u_{n -}^\delta + {}\\
  {} + (g_V + g_A) u^{\dag}_{e \alpha +} \gamma_{a \beta -}{}^\alpha
  u^\beta_{\nu +} u^{\dag}_{p \gamma +} \gamma^a_{\delta -}{}^\gamma
  u_{n +}^\delta  \biggr] = {} \\
  {}=\frac{4 G_F}{\sqrt{2}} \biggl[ (g_V -g_A) u^{\dag}_{e \alpha\,+}
  u^\alpha_{n\,-} u^{\dag}_{p \beta \,-}
  u^\beta_{\nu \, +} + {}  \\
  {} + (g_V + g_A) \left( u^{\dag}_{e \alpha +} u^\alpha_{\nu +}
    u^{\dag}_{p \beta +} u^\beta_{n +} - u^{\dag}_{e \alpha +}
    u^\alpha_{n +} u^{\dag}_{p \beta +} u^\beta_{\nu +} \right)
  \biggr].
\end{multline}

  Let us choose the direction angles as follows: $\varphi_\nu =
  \varphi_n = \varphi_p = \varphi_e = 0$, $\theta_\nu = \theta_n =
  \pi/2$, $\theta_e$ and $\theta_p$ are arbitrary.

  We introduce the spinors
\begin{equation}
  \label{eq:qed.matrix.example.4}
  \begin{gathered}
    |s_0\rangle =
    \begin{pmatrix}
      1 \\
      0
    \end{pmatrix}, \quad |s_1\rangle =
    \begin{pmatrix}
      0 \\
      1
    \end{pmatrix}, \\
    |s_2\rangle =
    \begin{pmatrix}
      \cos \theta_p/2 \\
      \sin \theta_p/2
    \end{pmatrix}, \quad |s_3\rangle =
    \begin{pmatrix}
      - \sin \theta_e/2 \\
      \cos \theta_e/2
    \end{pmatrix}.
  \end{gathered}
\end{equation}

  Then we can write (see \eqref{eq:qed.matrix.7})
\begin{gather}
  \label{eq:qed.matrix.example.5}
  u_{\nu\pm} = \frac{1}{\sqrt{2}} \left( \sqrt{E_\nu + m_\nu} \pm
    s_\nu \sqrt{E_\nu - m_\nu} \right) |s_0\rangle, \\
  u_{n\pm} = \frac{1}{\sqrt{2}} \left( \sqrt{E_n + m_n} \pm
    s_n \sqrt{E_n - m_n} \right) |s_1\rangle, \\
  u_{p\pm} = \frac{1}{\sqrt{2}} \left( \sqrt{E_p + m_p} \pm
    s_p \sqrt{E_p - m_p} \right) |s_2\rangle, \\
  u_{e\pm} = \frac{1}{\sqrt{2}} \left( \sqrt{E_e + m_e} \pm s_e
    \sqrt{E_e - m_e} \right) |s_3\rangle.
\end{gather}

  We obtain from \eqref{eq:qed.matrix.example.3}
\begin{multline}
  \label{eq:qed.matrix.example.6}
  M = \frac{G_F}{\sqrt{2}} \left( \sqrt{E_e + m_e} + s_e \sqrt{E_e -
      m_e} \right) 
  \times {} \\ {} \times
  \left( \sqrt{E_\nu + m_\nu} + s_\nu \sqrt{E_\nu -
      m_\nu} \right) 
  \times {} \\  {} \times 
  \Bigl[ (g_V - g_A) \left( \sqrt{E_n + m_n} - s_n \sqrt{E_n
      - m_n} \right) 
  \times {} \\ {} \times
  \left( \sqrt{E_p + m_p} - s_p \sqrt{E_p -
      m_p} \right)  \langle s_3|s_1\rangle \langle s_2|s_0\rangle 
  + {} \\  {} + 
  (g_V + g_A) \left( \sqrt{E_n + m_n} + s_n \sqrt{E_n - m_n}
  \right) \times {} \\ {} \times
  \left( \sqrt{E_p + m_p} + s_p \sqrt{E_p - m_p} \right)
  \times {} \\ {} \times \bigl( \langle s_3|s_0\rangle \langle
  s_2|s_1\rangle - \langle s_3|s_1\rangle \langle s_2|s_0\rangle
  \bigr)
  \Bigr] 
  = {} \\ {} = 
  \frac{G_F}{\sqrt{2}} \left( \sqrt{E_e + m_e} + s_e \sqrt{E_e -
      m_e} \right) 
  \times {} \\ {} \times
  \left( \sqrt{E_\nu + m_\nu} + s_\nu \sqrt{E_\nu -
      m_\nu} \right) 
  \times {} \\ {} \times 
  \Bigl[ (g_V - g_A) \left( \sqrt{E_n + m_n} - s_n \sqrt{E_n
      - m_n} \right) 
  \times {} \\ {} \times 
  \left( \sqrt{E_p + m_p} - s_p \sqrt{E_p -
      m_p} \right)  c\cos \theta_e/2 \cos \theta_p/2 - {} \\
  {} - (g_V + g_A)  \left( \sqrt{E_n + m_n} + s_n \sqrt{E_n - m_n}
  \right) 
  \times {} \\ {} \times 
  \left( \sqrt{E_p + m_p} + s_p \sqrt{E_p - m_p} \right)
  \times {} \\
  {} \times \bigl( \sin \theta_e/2 \cos \theta_p/2 + \cos \theta_e/2
  \cos \theta_p/2 \bigr) \Bigr].
\end{multline}

  Thus, the number of calculated terms is decreased considerably (the
  order $n$ instead of $n^2$); moreover, they have a rather simple
  form.

\section{Conclusions}
\label{sec:conclusion}

  \begin{enumerate}
  \item Semispinors are simpler objects than spinors.
  \item We propose using Lorentz two-spinors instead of Dirac
    four-spinors.
  \item In relativistic calculations, two-spinors seem more adequate
    than quaternions.
  \end{enumerate}

\bibliographystyle{abbrvnat}
  
\bibliography{bib/ref}

\begin{thebibliography}{2}
\providecommand{\natexlab}[1]{#1}
\providecommand{\url}[1]{\texttt{#1}}
\expandafter\ifx\csname urlstyle\endcsname\relax
  \providecommand{\doi}[1]{doi: #1}\else
  \providecommand{\doi}{doi: \begingroup \urlstyle{rm}\Url}\fi

\bibitem[Cartan(2012)]{cartan2012theory}
{\'E}.~Cartan.
\newblock \emph{The Theory of Spinors}.
\newblock Dover Books on Mathematics. Dover Publications, 2012.
\newblock ISBN 9780486137322.
\newblock URL \url{http://books.google.ru/books?id=AEZ1h7Cg3cwC}.

\bibitem[Penrose and Rindler(1987)]{penrose1987spinors}
R.~Penrose and W.~Rindler.
\newblock \emph{Spinors and Space-Time: Volume~1, Two-Spinor Calculus and
  Relativistic Fields}.
\newblock Cambridge Monographs on Mathematical Physics. Cambridge University
  Press, 1987.
\newblock ISBN 9780521337076.
\newblock URL \url{http://books.google.ru/books?id=CzhhKkf1xJUC}.

\end{thebibliography}


\begin{thebibliography}{1}
\def\selectlanguageifdefined#1{
\expandafter\ifx\csname date#1\endcsname\relax
\else\selectlanguage{#1}\fi}
\providecommand*{\href}[2]{{\small #2}}
\providecommand*{\url}[1]{{\small #1}}
\providecommand*{\BibUrl}[1]{\url{#1}}
\providecommand{\BibAnnote}[1]{}
\providecommand*{\BibEmph}[1]{#1}
\ProvideTextCommandDefault{\cyrdash}{\hbox to.8em{--\hss--}}
\providecommand*{\BibDash}{\ifdim\lastskip>0pt\unskip\nobreak\hskip.2em\fi
\cyrdash\hskip.2em\ignorespaces}

\bibitem{cartan:ru}
\selectlanguageifdefined{russian}
\BibEmph{Картан~Э.} Теория спиноров. \BibDash
\newblock Государственное издательство
  иностранной литературы, 1947. \BibDash
\newblock URL: \BibUrl{http://books.google.ru/books?id=zSwiOgAACAAJ}.

\bibitem{penrose1987spinors:ru}
\selectlanguageifdefined{russian}
\BibEmph{Пенроуз~Р., Риндлер~В.} Спиноры и
  пространство-время: Два-спинорное
  исчисление и релятивистские поля. Том~1:
  Пер. с англ. \BibDash
\newblock Мир, 1987.

\end{thebibliography}
  
\end{document}


\author{Д. С. Кулябов}
\email{yamadharma@gmail.com}
\author{А. Г. Ульянова}
\email{jelly26a@gmail.com}

\affiliation{Российский университет дружбы народов}

\thanks{Опубликовано в: Kulyabov~D.~S., Ul’yanova~A.~G. Application of
  two-spinor calculus in quantum mechanical and field calculations~//
  Physics of Particles and Nuclei Letters.~--- 2009.~--- Vol.~6,
  no.~7.~--- P.~546--549.~---
  \href{http://dx.doi.org/10.1134/S1547477109070115}{doi:~10.1134/S1547477109070115}.~---
  URL: \url{http://www.springerlink.com/content/bhmw776v15110271/}.}

\thanks{Исходные тексты:\\
  \url{https://bitbucket.org/yamadharma/articles-2008-two-spinors}}

\title{Применение 2-спинорного исчисления в квантовомеханических и
  полевых расчётах}

\begin{abstract}
  В статье описываются лоренцовы 2-спиноры и предлагается применять их
  в вычислениях, использующих дираковские 4-спиноры.
\end{abstract}

\maketitle

  \section{Введение}
\label{sec:intro}

  В физике достаточно интенсивно используются
  спиноры~\cite{cartan:ru}. Восновном применяются следующие спиноры:
  \begin{itemize}
  \item дираковские 4-спиноры;
  \item паулевские 3-спиноры;
  \item кватернионы.
  \end{itemize}

  При использовании дираковских 4-спиноров основную трудность вызывают
  $\gamma$-матрицы. По своей сути это объекты, служащие для связи
  спинорного и тензорного пространств и несущие, соответственно, два
  типа индексов: спинорные и тензорные. Логично было бы проводить
  вычисления лишь в одном из этих пространств.

  Предлагается использовать полуспиноры дираковских спиноров~---
  лоренцовы 2-спиноры~\cite{penrose1987spinors:ru} как более простые объекты.

  \section{Общее понятие спинора}
\label{sec:gener-noti-spin}

Определим спинор посредством уравнения Клиффорда-Дирака:
\begin{equation}
  \label{eq:1}
  \gamma_{(a}\gamma_{b)}=-g_{ab}\mathbf{I}.
\end{equation}

Или, опуская спинорные индексы:
\begin{equation}
  \label{eq:2}
  \gamma_{a\rho}^{\sigma} \gamma_{b\sigma}^{\tau} +
  \gamma_{b\rho}^{\sigma} \gamma_{a\sigma}^{\tau}= -2 g_{ab}\delta_{\rho}^{\tau}.
\end{equation}

Размерность спинорного пространства следующая:
\begin{equation}
  \label{eq:3}
    \begin{cases}
      N = 2^{n/2}, & \text{чётное } n; \\
      N = 2^{n/2 -1/2}, & \text{нечётное } n.
    \end{cases}
\end{equation}

\section{Связь 2-спиноров с кватернионами}
\label{sec:quat-two-spin}

  Положим
\begin{equation}
  \begin{gathered}
    \label{eq:quat.1}
    \mathbf{I} =
    \begin{pmatrix}
      1 & 0 \\
      0 & 1
    \end{pmatrix}, \quad \mathbf{i} =
    \begin{pmatrix}
      0 & \i \\
      \i & 0
    \end{pmatrix}, \\
    \mathbf{j} =
    \begin{pmatrix}
      0 & -1 \\
      1 & 0
    \end{pmatrix}, \quad \mathbf{k} =
    \begin{pmatrix}
      \i & 0 \\
      0 & -\i
    \end{pmatrix}.
  \end{gathered}
\end{equation}

  Общий кватернион будет представлен матрицей
\begin{equation}
  \label{eq:quat.2}
  \mathbf{A} = \mathbf{I} a + \mathbf{i} b + \mathbf{j} c + \mathbf{j} d =
  \begin{pmatrix}
    a + \i d & -c + \i b \\
    c + \i b & a - \i d
  \end{pmatrix},
\end{equation}
  где $a,b,c,d \in \setR$.
  Сумма и произведение двух кватернионов получаются просто как
  матричная сумма и матричное произведение.  Сопряжённый кватернион
  $\mathbf{A}^{*}$ определяется соответствующей матричной операцией:
\begin{equation}
  \label{eq:quat.3}
  \mathbf{A}^{*} = \mathbf{I} a - (\mathbf{i} b + \mathbf{j} c + \mathbf{k} d).
\end{equation}

  Если матрица $\mathbf{A}$ \eqref{eq:quat.2} унимодулярна и унитарна,
  то она будет представлять собой унитарную спин-матрицу. Можно
  записать соотношения:
\begin{gather}
  \label{eq:quat.4}
  \det \mathbf{A} = a^2 + b^2 + c^2 +d^2 = 1, \\
  \label{eq:quat.5}
  \mathbf{A}\mathbf{A}^{*} = \mathbf{I} (a^2 + b^2 +c^2 +d^2) =
  \mathbf{I}.
\end{gather}

  Таким образом, кватернион должен иметь единичную норму:
\begin{equation}
  \label{eq:quat.6}
  \| \mathbf{A} \| := a^2 + b^2 + c^2 +d^2 = 1.
\end{equation}
  Следовательно, единичный кватернион может быть представлен унитарной
  спин-матрицей.

  Несмотря на то, что унитарные спин-матрицы и единичные кватернионы
  представляют собой по существу одно и то же, в общем случае между
  кватернионами и спин-матрицами не существует такой тесной
  взаимосвязи. Дело в том, что кватернионы связаны с положительно
  определёнными квадратичными формами, тогда как спин"=матрицы и
  преобразования Лоренца характеризуются лоренцевой сигнатурой
  $(+,-,-,-)$.

\section{Связь 2-спиноров и векторов}
\label{sec:from-spinors-vectors}

При окончательных вычислениях возникает необходимость перевода
аббстрактных индексов в компонентную запись. Кроме того, зачастую
результат удобнее формулировать в векторной форме. Для установления
взаимосвязи между спинорным и векторным базисами служат символы
Инфельда--ван дер Вердена:
\begin{equation}
  \label{eq:spinor.verden.1}
  \begin{gathered}
    g_{\crd{a}}{}^{\crd{AA'}} := g_{\crd{a}}{}^a
    \varepsilon_A{}^{\crd{A}}
    \varepsilon_{A'}{}^{\crd{A'}}, \\
    g^{\crd{a}}{}_{\crd{AA'}} := g_a{}^{\crd{a}}
    \varepsilon_{\crd{A}}{}^A \varepsilon_{\crd{A'}}{}^{A'},
  \end{gathered}
\end{equation}
где свёртка производится только по абстрактным индексам.

Для стандартной тетрады Минковского и спиновой системы отсчёта
получим:
\begin{equation}
  \label{eq:spinor.verden.2}
  \begin{gathered}
    g_0{}^{\crd{AB'}} = \frac{1}{\sqrt{2}}
    \begin{pmatrix}
      1 & 0 \\
      0 & 1
    \end{pmatrix} = g_{\crd{AB'}}^0, \\ 
    g_1{}^{\crd{AB'}} =
    \frac{1}{\sqrt{2}}
    \begin{pmatrix}
      0 & 1 \\
      1 & 0
    \end{pmatrix} = g_{\crd{AB'}}^1, \\
    g_2{}^{\crd{AB'}} = \frac{1}{\sqrt{2}}
    \begin{pmatrix}
      0 & \i \\
      -\i & 0
    \end{pmatrix} = - g_{\crd{AB'}}^2, \\
    g_3{}^{\crd{AB'}} =
    \frac{1}{\sqrt{2}}
    \begin{pmatrix}
      1 & 0 \\
      0 & -1
    \end{pmatrix} = g_{\crd{AB'}}^3.
  \end{gathered}
\end{equation}

\section{Дираковские 4-спиноры и лоренцовы 2-спиноры}
\label{sec:dirac-four-spinors}

  В физических расчётах часто пользуются дираковскими
  4-спинорами\footnote{Введение 4-спиноров можно, наверное,
    мотивировать желанием сконструировать объект, с помощью которого
    можно было бы удобно реализовать операцию пространственного
    отражения.}.  Однако, операции с данными объектами крайне
  громоздки. Одним из главных недостатков этого формализма является
  явное использование $\gamma$-матриц, которые по сути своей являются
  объектами, служащими для связи векторного и спинорного
  пространств. Наши спинорные объекты <<живут>> как бы в двух
  пространствах: векторном и спинорном.  Соответственно наблюдается
  стремление перейти к более единообразному формализму посредством
  отказа либо от спинорных, либо от тензорных индексов.

  Рассмотрим переход к чисто спинорному формализму на основе
  лоренцовых 2-спиноров. Вначале сконструируем 4-спинорный объект на
  основе 2-спиноров. Далее продемонстрируем возможности данного
  формализма.  Рассмотрим два примера: вывод инвариантных спинорных
  соотношений и вычисление матричных элементов.

  \subsection{Построение 4-спинорного формализма}
\label{sec:qed.4spinor.build}

  Построим реализацию 4-спинорного формализма на основе лоренцовых
  2-спиноров\footnote{Зачастую дираковский 4-спинор реализуют с
    помощью двух трёхмерных спиноров (3-спиноров Паули). Формально это
    вполне возможно, поскольку структура пространств $\setS_R$ и
    $\setS_{R'}$ полуспиноров спинора размерности $n+1$ совпадает со
    структурой пространства $S_\rho$ спиноров размерности $n$ ($n$ ---
    нечётное)}.

  Будем обозначать строчными греческими буквами 4-спинорные
  индексы. Так же, как всегда, прописными латинскими буквами обозначим
  2-спинорные индексы, а строчными латинскими --- тензорные.

  Запишем дираковский 4-спинор как
\begin{equation}
  \label{eq:qed.4spinor.1}
  \psi^\alpha =
  \begin{pmatrix}
    \varphi^A \\
    \pi^{A'}
  \end{pmatrix},
\end{equation}
где $\varphi^A$ и $\pi^{A'}$ --- лоренцовы 2-спиноры.

  Сопряжённый спинор будет иметь вид:
\begin{equation}
  \label{eq:qed.4spinor.2}
  \overline{\psi^\alpha} = \bar{\psi}_\alpha =
  \begin{pmatrix}
    \bar \pi_A , \bar \varphi_{A'}
  \end{pmatrix}.
\end{equation}

  Определим оператор отражения:
\begin{equation}
  \label{eq:qed.4spinor.3}
  \hat P =
  \begin{pmatrix}
    \varphi^A \\
    \pi^{A'}
  \end{pmatrix} \mapsto
  \begin{pmatrix}
    \pi^{A'} \\
    \varphi^{A}
  \end{pmatrix}.
\end{equation}

  Будем использовать $\gamma$-матрицы в киральном представлении. Тогда
  запишем явный вид $\gamma$-матриц:
\begin{equation}
  \label{eq:qed.4spinor.4}
  \begin{gathered}
    \gamma_{a\rho}{}^\sigma = \sqrt{2}
    \begin{pmatrix}
      0 & \varepsilon_{A'R'} \varepsilon_{A}{}^{S} \\
      \varepsilon_{AR} \varepsilon_{A'}{}^{S'} & 0
    \end{pmatrix}, \\ 
    \eta_{\rho}{}^\sigma =
    \begin{pmatrix}
      -\i \varepsilon_{R}{}^S & 0 \\
      0 & \i \varepsilon_{R'}{}^{S'}
    \end{pmatrix},
  \end{gathered}
\end{equation}
  кроме того
\begin{equation}
  \label{eq:qed.4spinor.5}
  \gamma_{ab\rho}{}^\sigma =
  \begin{pmatrix}
    \varepsilon_{A'B'} \varepsilon_{R(A} \varepsilon_{B)}{}^S & 0 \\
    0 & \varepsilon_{AB} \varepsilon_{R'(A'} \varepsilon_{B')}{}^{S'}
  \end{pmatrix}.
\end{equation}

  Введём так же обозначение $\gamma_5 := \i \eta$.

  С помощью определённой нами структуры дираковских 4-спиноров можно
  построить инвариантные соотношения. Мы будем оперировать парой
  спинор--сопряжённый 4-спинор ($\psi$ и $\bar\psi$) и,
  соответственно, четырьмя 2-спинорами ($\varphi^A$,
  $\bar\varphi_{A'}$, $\pi^{A'}$ и $\bar\pi_A$).

  \subsection{Скаляры}
\label{sec:scalars}

  Свёртки $\bar\pi_A \varphi^A$ и $\bar\varphi_{A'}\pi_{A'}$ имеют
  смысл скаляров. Их сумма будет вести себя как скаляр, а разность как
  псевдоскаляр:
\begin{align}
  \label{eq:qed.4spinor.6}
  s &= \bar\pi_A \varphi^A + \bar\varphi_{A'}\pi^{A'}
  = \bar\psi_\alpha \psi^\alpha, \\
  \label{eq:qed.4spinor.7}
  p &= \i (\bar\pi_A \varphi^A - \bar\varphi_{A'}\pi^{A'}) = \i
  \bar\psi_\alpha \gamma_{5 \beta}{}^\alpha \psi^\beta.
\end{align}

  \subsection{Векторы}
\label{sec:vectors}

  Комбинации $\bar{\pi}^A\pi^{A'}$ и $\varphi^A\bar{\varphi}^{A'}$
  имеют смысл вектора. Их сумма ведёт себя как вектор, разность как
  псевдовектор:
\begin{align}
  \label{eq:qed.4spinor.8}
  j^a &= \sqrt{2} (\bar{\pi}^A\pi^{A'} + \varphi^A\bar{\varphi}^{A'})
  =
  \bar{\psi}_\alpha \gamma^{a}_ \beta{}^\alpha \psi^\beta,\\
  \label{eq:qed.4spinor.9}
  \tilde{j}^a &= \sqrt{2} (\bar{\pi}^A\pi^{A'} -
  \varphi^A\bar{\varphi}^{A'}) = \bar{\psi}_\alpha
  \gamma^{a}_\beta{}^\alpha \gamma_{5 \delta}{}^\beta \psi^\delta.
\end{align}

  \subsection{Тензоры}
\label{sec:tensors}

  Действительный кососимметричный тензор можно построить следующим
  образом:
\begin{equation}
  \label{eq:qed.4spinor.10}
  a^{ab} = \i (\varphi^{(A}\bar{\pi}^{B)} \varepsilon^{A'B'} -
  \bar{\varphi}^{(A'}\pi^{B')} \varepsilon^{AB}) = \bar{\psi}_\alpha
  \sigma^{ab}{}_\beta{}^\alpha \psi^\beta.
\end{equation}

  Эти соотношения нам представляются более простыми, нежели исходные.

  \section{Вычисление матричных элементов}
\label{sec:qed.matrix}

  Для расчёта матричных элементов в квантовой теории обычно применяют
  <<фейнмановский трюк>>, заключающийся в преобразовании произведения
  спиноров в след; в результате получают квадрат матричного элемента.
  Соответственно возрастает сложность вычислений --- количество
  вычисляемых членов пропорционально $n^2$. Кроме того, если полный
  матричный элемент вычисляется как сумма многих диаграмм, или важна
  информация о фазе, данный метод не применим.

  Как альтернатива предлагается вычислять сам матричный элемент.
  Рассмотрим два пути: применение 2-спиноров (отказ от тензорных
  индексов) и применение векторного формализма (отказ от спинорных
  индексов).

  Для ликвидации основного препятствия --- сложных соотношений для
  $\gamma$-матриц --- предлагается использовать 2-спинорный формализм.

  Введём вспомогательную нотацию, базирующуюся на знаке в проекторе $1
  \pm \gamma_5$:
\begin{equation}
  \label{eq:qed.matrix.1}
  \psi =
  \begin{pmatrix}
    \psi_{+} \\
    \psi_{-}
  \end{pmatrix}.
\end{equation}

  Соответственно $\gamma$-матрицы запишутся в виде (см.
  \eqref{eq:qed.4spinor.4}):
\begin{equation}
  \label{eq:qed.matrix.2}
  \gamma_a =
  \begin{pmatrix}
    0 & \gamma_{a\,+}\\
    \gamma_{a\,-} & 0
  \end{pmatrix}, \qquad \hat{p} =
  \begin{pmatrix}
    0 & \hat{p}_{+} \\
    \hat{p}_{-} & 0
  \end{pmatrix},
\end{equation}
  где $\hat{p}:= p^a \gamma_a$.

  Применение 2-спиноров особенно оправдано в случае наличия проекторов
  $(1 \pm \gamma_5)$. Поэтому рассмотрим упрощение вычислений в случае
  членов вида
\begin{equation}
  \label{eq:qed.matrix.3}
  \bar{\psi}_f \gamma^{a_1} \hat{p}_{(a)} \gamma^{a_2} \hat{p}_{(b)}
  \cdots \gamma^{a_n} \left[ \frac{1}{2} (1 \pm \gamma_5) \right] \psi_i.
\end{equation}

  Можно выделить два случая --- чётное и нечётное количество
  $\gamma$-матриц:
\begin{equation}
  \label{eq:qed.matrix.4}
  \begin{cases}
    \bar{\psi}_{f \pm} \gamma^{a_1}_{\mp} \hat{p}_{(a) \pm}
    \gamma^{a_2}_\mp \hat{p}_{(b) \pm} \cdots \gamma^{a_n}_\pm \psi_{i
      \pm}, \\
    \text{нечётное число $\gamma$-матриц;}
    \\
    \bar{\psi}_{f \mp} \gamma^{a_1}_{\pm} \hat{p}_{(a) \mp}
    \gamma^{a_2}_\pm \hat{p}_{(b) \mp} \cdots \gamma^{a_n}_\mp \psi_{i
      \pm}, \\
    \text{чётное число $\gamma$-матриц}.
  \end{cases}
\end{equation}

  Используя 2-спинорное представление $\gamma$-матриц, получаем
  следующие соотношения:
\begin{subequations}
  \label{eq:qed.matrix.5}
  \begin{gather}
    \label{eq:qed.matrix.5a}
    \gamma_{a\alpha}{}^\beta_{+} \gamma_{\gamma}^a{}^\delta_{-} = 2
    \delta_\alpha{}^\delta \delta_\gamma{}^\beta, \\
    \label{eq:qed.matrix.5b}
    \gamma_{a \alpha}{}^\beta_{\pm} \gamma_{\gamma}^a{}^\delta_{\pm} =
    2 \left( \delta_\alpha{}^\beta \delta_\gamma{}^\delta -
      \delta_\alpha{}^\delta \delta_\gamma{}^\beta \right).
  \end{gather}
\end{subequations}

  Для примера докажем \eqref{eq:qed.matrix.5a}:
\begin{equation}
  \label{eq:64}
  \varepsilon_{C'A'} \varepsilon_C{}^B \varepsilon^C{}_G
  \varepsilon^{C'D'} = 2 \varepsilon_{A'}{}^{D'} \varepsilon_G{}^B,
\end{equation}

  Учитывая соответствия
\begin{equation}
  \alpha \leftrightarrow A', \quad \beta \leftrightarrow B, \quad
  \gamma \leftrightarrow G, \quad \delta \leftrightarrow D',
\end{equation}
  получаем исходное выражение.

  Таким образом, применяя последовательно \eqref{eq:qed.matrix.4} и
  \eqref{eq:qed.matrix.5} мы освобождаемся от $\gamma$-матриц и
  производим вычисления с 2-спинорами.

  После вычислений получаются члены типа
\begin{equation}
  \label{eq:qed.matrix.6}
  u_{f\pm}^\dag \hat{p}_{(a)\mp} \hat{p}_{(b)\pm} \dots \hat{e}_{+\text{
      or }-} \dots u_{i\pm},
\end{equation}
  где $e$ --- поляризация.

  Для получения конкретного результата необходимо выбрать
  представление спинора. Например, можно использовать стандартное
  решение, полученное разложением по плоским волнам.  Тогда в случае
  продольной поляризации получим:
\begin{equation}
  \label{eq:qed.matrix.7}
  u_{\pm} = \left( \sqrt{E+\varepsilon m} \pm \varepsilon s \sqrt{E-
      \varepsilon m} \right)
  \begin{pmatrix}
    \e^{-\i \varphi/2} \sqrt{1 + s \cos \theta} \\
    \e^{\i \varphi/2} \sqrt{1 - s \cos \theta}
  \end{pmatrix},
\end{equation}
  где $s$ --- спиральность, $\varepsilon$ --- знак энергии.

  \subsection{Пример вычисления матричного элемента}
\label{sec:qed.matrix.example}

  Вычислим сечение реакции
\begin{equation}
  \label{eq:qed.matrix.example.1}
  \nu + n \to p + e^{-},
\end{equation}

  Матричный элемент которой в стандартной (V--A) теории имеет вид
\begin{equation}
  \label{eq:qed.matrix.example.2}
  M = \frac{G_F}{\sqrt{2}} \left( \bar{\psi}_e \gamma_a (1 + \gamma_5)
    \psi_\nu \right) \left( \bar{\psi}_p \gamma^a (g_V + g_A \gamma_5)
    \psi_n \right).
\end{equation}

  Используя \eqref{eq:qed.matrix.4} и \eqref{eq:qed.matrix.5} приведём
  \eqref{eq:qed.matrix.example.2} к виду:
\begin{multline}
  \label{eq:qed.matrix.example.3}
  M = \frac{2 G_F}{\sqrt{2}} \bigl(u^{\dag}_{e\alpha +} \gamma_{a
    \beta -}{}^\alpha u_{\nu +}^\beta \bigr) 
  \times {} \\ {} \times
  \biggl[ (g_A - g_V)
  \bigl( u^{\dag}_{p \gamma +} \gamma^a_{\delta -}{}^\gamma
  u^\delta_{n +} + u^{\dag}_{p \gamma -} \gamma^a_{\delta +}{}^\gamma
  u^\delta_{n -} \bigr) + {} \\
  {} + 2 g_A u^{\dag}_{p \gamma +} \gamma^a_{\delta -}{}^\gamma
  u^\delta_{n +} \biggr] 
  = {} \\ {} =
  \frac{2 G_F}{\sqrt{2}} \biggl[ (g_V - g_A)
  u^{\dag}_{e \alpha +} \gamma_{a \beta -}{}^\alpha u^\beta_{\nu +}
  u^{\dag}_{p \gamma -} \gamma^a_{\delta +}{}^\gamma u_{n -}^\delta + {}\\
  {} + (g_V + g_A) u^{\dag}_{e \alpha +} \gamma_{a \beta -}{}^\alpha
  u^\beta_{\nu +} u^{\dag}_{p \gamma +} \gamma^a_{\delta -}{}^\gamma
  u_{n +}^\delta  \biggr] = {} \\
  {}=\frac{4 G_F}{\sqrt{2}} \biggl[ (g_V -g_A) u^{\dag}_{e \alpha\,+}
  u^\alpha_{n\,-} u^{\dag}_{p \beta \,-}
  u^\beta_{\nu \, +} + {}  \\
  {} + (g_V + g_A) \left( u^{\dag}_{e \alpha +} u^\alpha_{\nu +}
    u^{\dag}_{p \beta +} u^\beta_{n +} - u^{\dag}_{e \alpha +}
    u^\alpha_{n +} u^{\dag}_{p \beta +} u^\beta_{\nu +} \right)
  \biggr].
\end{multline}

  Выберем направляющие углы следующим образом: $\varphi_\nu =
  \varphi_n = \varphi_p = \varphi_e = 0$, $\theta_\nu = \theta_n =
  \pi/2$, $\theta_e$ и $\theta_p$ --- произвольные.

  Введём спиноры:
\begin{equation}
  \label{eq:qed.matrix.example.4}
  \begin{gathered}
    |s_0\rangle =
    \begin{pmatrix}
      1 \\
      0
    \end{pmatrix}, \quad |s_1\rangle =
    \begin{pmatrix}
      0 \\
      1
    \end{pmatrix}, \\
    |s_2\rangle =
    \begin{pmatrix}
      \cos \theta_p/2 \\
      \sin \theta_p/2
    \end{pmatrix}, \quad |s_3\rangle =
    \begin{pmatrix}
      - \sin \theta_e/2 \\
      \cos \theta_e/2
    \end{pmatrix}.
  \end{gathered}
\end{equation}

  Тогда можно записать (см. \eqref{eq:qed.matrix.7}):
\begin{gather}
  \label{eq:qed.matrix.example.5}
  u_{\nu\pm} = \frac{1}{\sqrt{2}} \left( \sqrt{E_\nu + m_\nu} \pm
    s_\nu \sqrt{E_\nu - m_\nu} \right) |s_0\rangle, \\
  u_{n\pm} = \frac{1}{\sqrt{2}} \left( \sqrt{E_n + m_n} \pm
    s_n \sqrt{E_n - m_n} \right) |s_1\rangle, \\
  u_{p\pm} = \frac{1}{\sqrt{2}} \left( \sqrt{E_p + m_p} \pm
    s_p \sqrt{E_p - m_p} \right) |s_2\rangle, \\
  u_{e\pm} = \frac{1}{\sqrt{2}} \left( \sqrt{E_e + m_e} \pm s_e
    \sqrt{E_e - m_e} \right) |s_3\rangle.
\end{gather}

  Из \eqref{eq:qed.matrix.example.3} получаем
\begin{multline}
  \label{eq:qed.matrix.example.6}
  M = \frac{G_F}{\sqrt{2}} \left( \sqrt{E_e + m_e} + s_e \sqrt{E_e -
      m_e} \right) 
  \times {} \\ {} \times
  \left( \sqrt{E_\nu + m_\nu} + s_\nu \sqrt{E_\nu -
      m_\nu} \right) 
  \times {} \\  {} \times 
  \Bigl[ (g_V - g_A) \left( \sqrt{E_n + m_n} - s_n \sqrt{E_n
      - m_n} \right) 
  \times {} \\ {} \times
  \left( \sqrt{E_p + m_p} - s_p \sqrt{E_p -
      m_p} \right)  \langle s_3|s_1\rangle \langle s_2|s_0\rangle 
  + {} \\  {} + 
  (g_V + g_A) \left( \sqrt{E_n + m_n} + s_n \sqrt{E_n - m_n}
  \right) \times {} \\ {} \times
  \left( \sqrt{E_p + m_p} + s_p \sqrt{E_p - m_p} \right)
  \times {} \\ {} \times \bigl( \langle s_3|s_0\rangle \langle
  s_2|s_1\rangle - \langle s_3|s_1\rangle \langle s_2|s_0\rangle
  \bigr)
  \Bigr] 
  = {} \\ {} = 
  \frac{G_F}{\sqrt{2}} \left( \sqrt{E_e + m_e} + s_e \sqrt{E_e -
      m_e} \right) 
  \times {} \\ {} \times
  \left( \sqrt{E_\nu + m_\nu} + s_\nu \sqrt{E_\nu -
      m_\nu} \right) 
  \times {} \\ {} \times 
  \Bigl[ (g_V - g_A) \left( \sqrt{E_n + m_n} - s_n \sqrt{E_n
      - m_n} \right) 
  \times {} \\ {} \times 
  \left( \sqrt{E_p + m_p} - s_p \sqrt{E_p -
      m_p} \right)  c\cos \theta_e/2 \cos \theta_p/2 - {} \\
  {} - (g_V + g_A)  \left( \sqrt{E_n + m_n} + s_n \sqrt{E_n - m_n}
  \right) 
  \times {} \\ {} \times 
  \left( \sqrt{E_p + m_p} + s_p \sqrt{E_p - m_p} \right)
  \times {} \\
  {} \times \bigl( \sin \theta_e/2 \cos \theta_p/2 + \cos \theta_e/2
  \cos \theta_p/2 \bigr) \Bigr].
\end{multline}

  Таким образом мы получили существенное (порядок: $n$ вместо $n^2$)
  сокращения количества вычисляемых членов, кроме того, все они имеют
  достаточно простой вид.

\section{Заключение}
\label{sec:conclusion}

  \begin{enumerate}
  \item Полуспиноры более простые объекты, чем спиноры.
  \item Предлагается использовать лоренцовы 2-спиноры вместо
    дираковских 4-спиноров.
  \item При релятивистских расчётах аппарат 2-спиноров представляется
    более адекватным, чем аппарат кватернионов.
  \end{enumerate}

\bibliographystyle{gost2008l}
  
\bibliography{bib/ref}